\documentclass[prb,twocolumn,showpacs,amsmath,nobibnotes,nofootinbib,floatfix]{revtex4}
\usepackage[latin1]{inputenc}
\usepackage{color}
\usepackage{ulem}
\usepackage{graphicx}
\usepackage{epstopdf}
\usepackage{bm}
\usepackage{amssymb}
\usepackage{amsmath}

\begin{document}
\title{Ratchet effects in two-dimensional systems with a lateral periodic potential}

\author{A.~V.~Nalitov}
\author{L.~E.~Golub}
\email{golub@coherent.ioffe.ru}
\author{E.~L.~Ivchenko}
\affiliation{Ioffe
Physical-Technical Institute of the Russian Academy of Sciences, 194021
St.~Petersburg, Russia}

\begin{abstract}
Radiation-induced ratchet electric currents have been studied theoretically in graphene with a periodic noncentrosymmetric lateral potential. The ratchet current generated under normal incidence is shown to consist of two contributions, one of them being polarization-independent and proportional to the energy relaxation time, and another controlled solely by 
elastic scattering processes and sensitive to both the linear and circular polarization of radiation. Two realistic mechanisms of electron scattering in graphene are considered. For short-range defects, the ratchet current is helicity-dependent but independent of the direction of linear polarization. For the Coulomb impurity scattering, the ratchet current is forbidden for the radiation linearly polarized in the plane perpendicular to
the lateral-potential modulation direction. 
For comparison, the ratchet currents in a quantum well with a lateral superlattice are calculated at low temperatures with allowance for 
the dependence of the momentum relaxation time on the electron energy.
\end{abstract}

\pacs{
68.65.Pq,        %Graphene films
72.80.Vp,  %Electronic transport in graphene
65.80.Ck,       % Thermal properties of graphene
73.63.Hs        
%73.63.-b        Electronic transport in nanoscale materials and structures (see  
%also 73.23.-b Electronic transport in mesoscopic systems)
%Quantum wells
}

%\date{\today}

\maketitle

\section{Introduction}\label{sec:intro}
Noncentrosymmetric periodic systems being driven out of thermal equilibrium by a time-oscillating force, stochastic or deterministic, are able to transport particles even if the force is zero on average. This directed transport, generally known as the ratchet effect, is relevant to different fields of natural sciences. Various kinds of symmetry-breaking micro- and nanometer-sized artificial structures have been proposed and fabricated to model ratchets and investigate their fundamental properties, 
for review see, e.g., Refs.~\onlinecite{reimann,motorsAppl,demich,Review_JETP_Lett}. 
Directed motion of Brownian particles in water have been induced by modulating in time a spatially periodic but asymmetric optical potential.~\cite{water}
Electronic ratchets where rectification of thermal fluctuations is achieved in systems with inhomogeneous distribution of temperature have been studied 
theoretically in Refs.~\onlinecite{buetiker1,buetiker2,Buttiker_2012}. 
In semiconductor nanostructures, the ratchet effect has been demonstrated in various systems
with asymmetric scatterer arrays based on both A$_3$B$_5$~\cite{kotthaus,samuelson,Chepel,Kvon} and Si/Ge~\cite{Kannan} materials as well as 
with asymmetric lateral superlattices.~\cite{wegscheider,popov} 
The spin ratchets have been proposed for two-dimensional systems with symmetric  periodic potential and driving force but with the Rashba spin-orbit interaction.~\cite{richter,grifoni}
Recently, the ratchet current induced by terahertz radiation has been observed in semiconductor quantum-well (QW) structure with a one-dimensional lateral periodic potential induced either by etching a noncentrosymmetric grating into the sample cap layer\cite{Olbrich_PRL_09} or by deposition of micropatterned metal-gate fingers.\cite{Olbrich_PRB_11}
Quantum graphene ratchets formed by asymmetric periodic strain with the period comparable with the de~Broglie wavelength of free carriers have been studied in Ref.~\onlinecite{KiselevGolub}. 

\begin{figure}[t]
\includegraphics[width=0.8\linewidth]{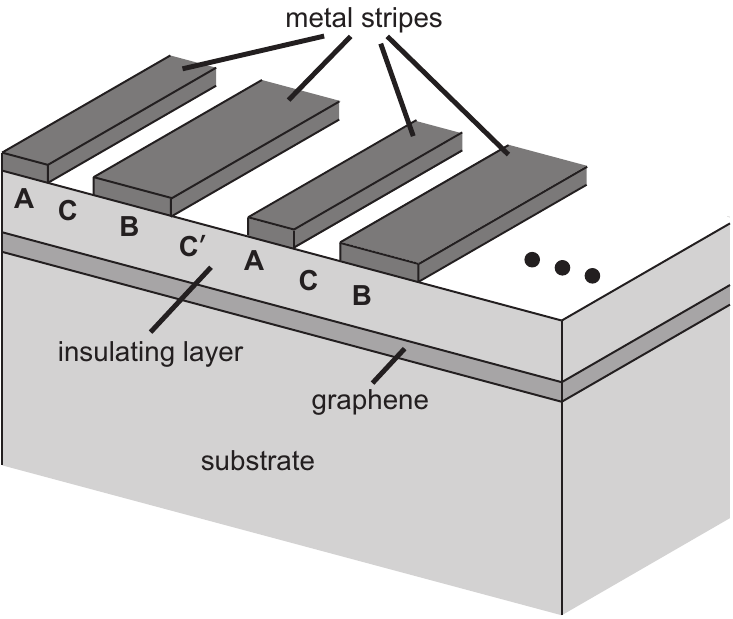}
\caption{Schematic representation of the studied ratchet structure.}
\label{fig_structure}
\end{figure}

In the present theoretical work we consider a classical graphene ratchet consisting of a graphene sheet fabricated on highly resistive substrate and covered successively  by a thin dielectric layer and a periodic grating of semi-transparent metal fingers, Fig.~\ref{fig_structure}. 
A technological opportunity to grow such a system is demonstrated in Refs.~\onlinecite{gr_photodetect,gr_plasmonic_struct} where photodetectors with multiple interdigitated metal fingers fabricated on the graphene were reported.
To achieve symmetry breaking the superimposed lateral structure may form the infinite sequence 
$...ACBC'ACBC'...$ with $A$, $B$ representing metal fingers of the thicknesses $a$ and $b$ ($a \neq b$), and
$C$, $C'$ representing hollows of different thicknesses $c$ and $c'$, Fig.~\ref{fig_structure}. 
%
%To achieve symmetry breaking the superimposed lateral structure may form the infinite sequence $...ACBCACBC...$ with $A$, $B$ representing metal fingers of different thicknesses, and
%$C$ representing hollows in between, Fig.~\ref{fig_structure}. 

The paper is organized as follows. In Sec.~\ref{sec:Basic} we formulate the basic concept of the problem. In Sec.~\ref{sec:Seebeck} we take into account the radiation-induced heating of the free carriers and spatial modulation of the heating, 
and deduce the polarization-independent contribution to the ratchet current.
In Sec.~\ref{sec:kinetic} we develop the Boltzmann kinetic formalism and obtain microscopic expressions for the polarization-dependent ratchet currents. The results are discussed in Sec.~\ref{sec:discussion}. In Sec.~\ref{sec:concl} the summary of the research is outlined.

\section{Basic concept} 
\label{sec:Basic}

According to the theory proposed in Refs.~\onlinecite{Olbrich_PRL_09,Olbrich_PRB_11} in the studied systems, the pulsating ratchets, the electric current generation is based on the combined action of a static spatially-periodic in-plane potential 
\[
V(x) =V_0 \cos{(qx + \varphi_V)}
\] 
and a spatially modulated electric-field amplitude 
\[
{\bm E}(x) = {\bm E}_0 [1 + h \cos{(qx + \varphi_E)}]
\]
of the normally-incident radiation. Here 
$q = 2 \pi/d$ with $d$ being the superlattice period along $x$. In the considered pulsating ratchets, the symmetry-breaking is described by the phase shift $\varphi_V - \varphi_E$ different from an integer number of $\pi$, and the electromotive force is proportional to $\sin{(\varphi_V - \varphi_E)}$. The structured graphene presented in Fig.~\ref{fig_structure} can also show ratchet effects. 
The lateral potential $V(x)$ can arise due to the strain, tensile or compressive, in graphene areas located beneath the fingers $A$ and $B$ whereas the in-plane modulation of the pump radiation appears due to near-field effects of the THz radiation propagating through the grating. For $A\neq B$, $C\neq C'$, the shapes and local extrema of the periodic functions $V(x)$ and $\bm E(x)$ 
are naturally shifted with respect to each other resulting in a difference between the phases $\varphi_V$ and
$\varphi_E$. 
For example, for $b=c=2a=2c'$, a crude estimate gives a value of 0.4 for $\sin{(\varphi_V - \varphi_E)}$.

The symmetry imposes restrictions on the polarization dependence of the
ratchet currents. Graphene modulated by an asymmetric lateral potential $V(x)$ has the
point group symmetry $C_{2v}$ with the $C_2$ axis parallel to the modulation direction $x$ and the mirror reflection plane
perpendicular to $y$. It follows then that the net dc current density components $j_x$ and $j_y$ are related to components of the polarization unit vector $\bm e$ and amplitude $E(x) \equiv |{\bm E}(x)|$ of the normally incident radiation by
four linearly independent coefficients
\begin{eqnarray} 
&&j_x=\left[ \chi_0 (\vert e_x \vert^2 + \vert e_y \vert^2) + \chi_L (\vert e_x \vert^2 - \vert e_y \vert^2) \right] \overline{E^2(x) {dV \over dx}}\:, \nonumber\\
&&j_y=\left[ \tilde{\chi}_L (e_x e_y^*+ e_x^* e_y) + \gamma P_{\rm circ} \right] \overline{E^2(x) {dV \over dx}}\:. \label{phenom}
\end{eqnarray}
Here 
we use the notations 
$x$ and $y$ for the in-plane coordinates,
the bar denotes averaging over the coordinate $x$, 
and $P_{\rm circ} = {\rm i} (e_x e_y^*- e_x^* e_y)$ is the degree of the radiation circular polarization. The coefficient $\chi_0$ describes the contribution to the ratchet current insensitive to the polarization state, 
while the remaining three coefficients describe the linear ($\chi_L, \tilde{\chi}_L$) and circular ($\gamma$) ratchet effects.
We develop the kinetic theory allowing us to derive equations for all these coefficients. As compared to Refs.~\onlinecite{Olbrich_PRL_09,Olbrich_PRB_11} the theory will be substantially extended to take into account specific properties of graphene, namely, (i)~the linear, Dirac-like dispersion of electron energy leads to a strong energy dependence of elastic relaxation times and (ii)~the electron gas in doped graphene is degenerate even at room temperature.

Taking the electric-field and lateral-potential modulation in the above simplest form
we obtain for the average in Eq.~(\ref{phenom})
\begin{equation} \label{E0V}
\overline{E^2(x) {dV \over dx}} = q V_0 h E_0^2 \sin{(\varphi_V - \varphi_E)}\:.
\end{equation}
It is worth noting that the factor $h \sin{(\varphi_V - \varphi_E)}$ in Eq.~(\ref{E0V}) 
should be strongly sensitive to the geometry of the structured coating of the graphene, namely, the thicknesses of stripes and heights of the metal fingers. Furthermore, bias voltages applied to the grid elements $A$ and $B$ should substantially change the lateral potential $V(x)$ as well as the relative amplitude $h$ and phase $\varphi_E$ of the field modulation. The replacement of thin semi-transparent metal fingers by thick gate stripes should reveal plasmonic effects similarly to photoresponse of the THz plasmonic broadband detectors, see Refs.~\onlinecite{popov,Drexler} and references therein.

Hereafter we consider a graphene sheet with the lateral potential $V(x)$. 
The electron energy in each valley, $K$ or $K'$, is given by
\begin{equation} \label{energy}
E_{\bm k} = \hbar v_0 k + V(x)\:,
\end{equation}
where $v_0$ is the electron speed in graphene and the two-dimensional wave vector ${\bm k}$ is referred to the vortex of the hexagonal Brillouin zone. Since in the model under consideration the behavior of electrons in the $K$ or $K'$ valleys is identical we consider the current generation in one of them and then double the result. In the course of presenting the results we will supplement them with similar results obtained for the QW ratchet with the electron parabolic dispersion $E_{\bm k} = \hbar^2 k^2/2 m + V(x)$. For adequate comparison of the two low-dimensional systems we have extended the  theory of Ref.~\onlinecite{Review_JETP_Lett} for QW ratchets to take into account the dependence of the momentum relaxation time on the electron energy and the degenerate statistics.

\section{Seebeck ratchet current} 
\label{sec:Seebeck}

In this and next sections we will successively consider two mechanisms of the ratchet current. 
In the Seebeck ratchet effect, the spatially-modulated radiation heats the electron gas changing its effective temperature 
from the equilibrium value $T$ to $T(x) = \bar{T} + \delta T(x)$. Here $\bar{T}$ is the average electron temperature and 
$\delta T(x)$ oscillates in space with the period $d$. In turn, the correction $\delta T(x)$ causes the inhomogeneous
correction to the conductivity $\delta \sigma(x) = (\partial \sigma/ \partial T) \delta T(x)$. Bearing in mind Ohm's law
${\bm j} = \sigma {\bm E}$, replacing the dc electric field ${\bm E}$ by $- (1/e) dV/dx$ and $\sigma$ by $\delta \sigma(x)$  
we obtain for the ratchet current
\begin{equation} \label{j_S}
j_x = {1\over |e|} \overline{ \delta\sigma(x) \frac{dV(x)}{dx}}\:.
\end{equation}
Here $e<0$ is the electron charge.
The nonequilibrium electron temperature can be found
from the energy balance equation
\begin{equation} \label{balance}
{T(x)-T\over \tau_\varepsilon} = \hbar \omega G(x)\:,
\end{equation}
where $\tau_{\varepsilon}$ is the electron energy relaxation time, $\omega$ is the radiation frequency, the temperature is expressed in energy units, and $G(x)$ is the photon absorption rate per electron. From Eqs.~(\ref{j_S}), (\ref{balance}) we derive the working equation
\begin{equation} \label{seebeck}
j_x = \frac{\hbar \omega \tau_{\varepsilon}}{|e|} \frac{\partial \sigma}{\partial T} \overline{G(x) \frac{dV(x)}{dx}}\:,
\end{equation}
which can be used for a degenerate two-dimensional gas in both graphene and QWs. In graphene, the Drude absorption rate per particle is inversely proportional to the Fermi energy $\varepsilon_{\rm F}$,
\begin{equation} \label{rate_graphene}
G(x) = {e^2 v_0^2 \over \varepsilon_{\rm F}} { \tau_{\rm tr} \over 1 + (\omega  \tau_{\rm tr})^2} {2 E^2(x)\over \hbar \omega},
\end{equation}
while for QWs $v_0^2/\varepsilon_{\rm F}$ should be replaced by the inversed electron effective mass $1/m$. Here $\tau_{\rm tr}$ is the transport relaxation time which determines the low-temperature conductivity.

The temperature dependence of conductivity for degenerate electron gas in graphene is well documented~\cite{Das_Sarma2009}
\begin{equation} \label{sigmagr}
\frac{\partial \sigma}{\partial T} = {\pi e^2\over 3 \hbar^2} T \varepsilon_{\rm F} \left[  {(\varepsilon\tau_1)' \over \varepsilon} \right]'_{\varepsilon=\varepsilon_{\rm F}}\:,
\end{equation}
while for the QW systems one has
\begin{equation} \label{ds_T_QW}
\frac{\partial \sigma_{\rm QW}}{\partial T} = {\pi e^2\over 3\hbar^2} T (\varepsilon\tau_1)^{\prime\prime}_{\varepsilon=\varepsilon_{\rm F}}\:.
\end{equation}
Here primes denote differentiation over the electron energy $\varepsilon \equiv \varepsilon_k = \hbar v_0 k$ or $\hbar^2k^2/(2m)$, and $\tau_1(\varepsilon)$ is the momentum relaxation time of a nonequilibrium correction to the electron distribution function depending as $\cos{\varphi_{\bm k}}$ on the azimuthal angle $\varphi_{\bm k}$ of the electron wave vector ${\bm k}$. Note that the transport relaxation time $\tau_{\rm tr}$ is equal to $\tau_1(\varepsilon_{\rm F})$.

From Eqs.~\eqref{seebeck}-\eqref{ds_T_QW} we finally obtain the Seebeck contribution to the polarization-independent current Eq.~\eqref{phenom} described by coefficient $\chi_0$. For graphene it is given by
\begin{equation}
\label{chi_Seebeck}
\chi_{0}^{\rm S} = -{2 \pi e^3  v_0^2 {T} \tau_\varepsilon \over 3\hbar^2 \varepsilon_{\rm F}} { \tau_{\rm tr} \over 1 + (\omega  \tau_{\rm tr})^2}  \varepsilon_{\rm F} \left[  {(\varepsilon\tau_1)' \over \varepsilon} \right]'_{\varepsilon=\varepsilon_{\rm F}}\:,
\end{equation}
while for the QW structures one has
\begin{equation} \label{chi_Seebeck_QW}
	\chi_{0, {\rm QW}}^{\rm S} = -{2 \pi e^3  {T} \tau_\varepsilon \over 3\hbar^2 m} { \tau_{\rm tr} \over 1 + (\omega  \tau_{\rm tr})^2}  \left(\varepsilon \tau_1\right)^{\prime\prime}_{\varepsilon=\varepsilon_{\rm F}}\:.
\end{equation}
The analysis of Eqs.~\eqref{chi_Seebeck} and~\eqref{chi_Seebeck_QW} for different mechanisms of electron scattering is postponed to Sec.~\ref{sec:discussion}.

\section{Polarization-dependent ratchet currents} \label{sec:kinetic}
In the presence of normally-incident radiation, an electron is subjected to the periodic force
\begin{equation} \label{force}
{\bm F}(x) = e \left[ {\bm E}(x) {\rm e}^{- {\rm i} \omega t} + {\rm c.c.} \right] - {dV(x) \over dx} \, \hat{\bm x}\:,
\end{equation}
where $\hat{\bm x}$ is the unit vector in the $x$ direction.
In the previous section we could avoid the consideration in terms of the Boltzmann kinetic equation for the electron distribution function $f_{\bm k}$ and, instead, used the known expressions for the electron conductivity. However, the second mechanism of the ratchet currents should be treated on the base of the Boltzmann equation
\begin{equation} \label{Boltzmann}
\left( \frac{\partial}{\partial t} + v_{\bm k,x}
\frac{\partial}{\partial x} + \frac{{\bm F}(x)}{\hbar}
\frac{\partial}{\partial {\bm k}} \right)
f_{\bm k}(x) + Q_{\bm k}(f)  = 0\:.
\end{equation} 
Here ${\bm v}_{\bm k}$ is the velocity of an electron with the wavevector ${\bm k}$ equal to $v_0 {\bm k}/k$ in graphene and
$\hbar {\bm k}/m$ in a conventional QW, and $Q_{\bm k}$ is the collision integral. In what follows we assume that
$\tau_{\rm tr}, \omega^{-1} \ll \tau_{\varepsilon}$ 
and neglect the energy relaxation in Eq.~(\ref{Boltzmann}) in which case the integral $Q_{\bm k}$ describes only momentum relaxation processes. Thus, in the second mechanism the ratchet current can be independent of the energy relaxation time whereas in the first mechanism related to the carrier heating the current is proportional to $\tau_{\varepsilon}$.

In terms of the distribution function the electric current density in graphene is written as 
\begin{equation} \label{j}
{\bm j} = \nu e \sum\limits_{\bm k} {\bm v}_{\bm k} \overline{f_{\bm k}(x)}\:,
\end{equation}
where the factor $\nu$ accounts for the spin and valley degeneracy, in QW structures $\nu = 2$ and in graphene $\nu =4$. 

In order not to overload the theory with too cumbersome equations we impose the following properties of the system under consideration: 
the electron mean free path $l_e = v_0 \tau_{\rm tr}$ and energy diffusion length
$l_{\varepsilon} = v_0 \sqrt{\tau_{\rm tr} \tau_{\varepsilon}}$ 
are both small compared with the superlattice period $d$; and the ac diffusion is neglected which is valid if $v_0 \ll \omega d$. On the other hand, no restrictions are imposed on the value of $\omega \tau_{\rm tr}$. Moreover, we assume the radiation electric field and the lateral potential to be weak enough, so that
\[
|e E_0| v_0 \tau_{\rm tr} \ll T
\quad
\mbox{and} 
\quad
\left|V(x) - \overline{V(x)}\right| \ll \varepsilon_{\rm F}\:.
\]
Then, according to the phenomenological Eqs.~\eqref{phenom} the function $\overline{f_{\bm k}(x)}$ should be calculated in the third order of the perturbation theory, including the second order in the electric-field amplitude and the first order in the lateral potential.  Taking into account that our aim is to derive 
expression for the sum (\ref{j}) rather than to find the function $f_{\bm k}(x)$ explicitly we can express this sum in the form
\begin{equation} \label{j1}
{j_{\alpha}} = {\nu  e^2  \over \hbar} \sum_{\bm k}  \overline{ f^{(EV)}_{{\bm k} \omega}(x) {\bm E}^*(x)} \cdot  \frac{\partial \bigl(\tau_1 v_{\alpha}  \bigr)}{\partial \bm k}  + {\rm c.c.}\:,
\end{equation}
where $\alpha = x,y$, and $f^{(EV)}_{{\bm k} \omega}$ is the second-order iteration linear both in ${\bm E}(x)$ and $dV(x)/dx$. It can be found from the equation
\begin{eqnarray} \label{EV}
\left( - {\rm i} \omega + v_{\bm k,x}
\frac{\partial}{\partial x} \right) f^{(EV)}_{{\bm k}\omega}(x)
+  Q_{\bm k} \bigl(f^{(EV)}\bigr)
\nonumber
\\ = -\frac{e}{\hbar}\ {\bm E}(x) {\partial f^{(V)} \over \partial \bm k} + {dV \over dx} {1 \over \hbar} {\partial f^{(E)} \over \partial k_x}\:, \hspace{0.4 cm}
\end{eqnarray}
where the first-order corrections are given by
\begin{equation}
f^{(V)}_k=V(x) f_0'(\varepsilon_k), \quad
f^{(E)}_{\bm k}=-e \tau_{1\omega} \bm E(x) \bm v_{\bm k} f_0'(\varepsilon_k)\:.
\end{equation}
Here $f_0(\varepsilon_k)$ is the equilibrium Fermi-Dirac function at $V(x) \equiv 0$, and
$\tau_{1\omega} = \tau_{1}/(1 - {\rm i} \omega \tau_1)$.

\subsection{Graphene}
In contrast to 
%QW structures 
systems
with the parabolic energy dispersion, in graphene the relaxation time $\tau_1$ is energy-dependent even for the short-range scattering potential. Moreover, the derivative $\partial v_{\alpha}/ \partial k_{\beta}$ is ${\bm k}$-dependent which means that the current~\eqref{j1} cannot be in general expressed exclusively in terms of the macroscopic fluctuation $\delta N_\omega(x) = 4 \sum\limits_{\bm k} f^{(EV)}_{{\bm k} \omega}$. 

To calculate the ratchet current, we solve the kinetic Eq.~\eqref{EV} and find a contribution to $f^{(EV)}_{{\bm k}\omega}(x)$ even in ${\bm k}$. It is convenient to present it as a sum of isotropic part $\left< f^{(EV)}_{k \omega} (x) \right>$ and anisotropic part $\delta f^{(EV)}_{{\bm k} \omega}(x)$, where the angular brackets denote averaging over the directions of ${\bm k}$. The isotropic part describing a nonequilibrium correction to the energy distribution of electrons has the form
\begin{eqnarray} \label{f_EV_0} 
\left< f^{(EV)}_{k \omega} \right> = {{\rm i} e v_0^2\over 2 \omega} \hspace{3 cm}\\ \times
\left[ (-f_0') {\left( \varepsilon_k \tau_{1 \omega} \right)' \over \varepsilon_k} E_{x}(x) {dV\over dx}
+ f_0^{\prime\prime} \tau_{1 \omega} V(x ) {d E_{x}\over dx} 	\right]\:. \nonumber
\end{eqnarray}
In fact, Eq.~(\ref{f_EV_0}) describes the local oscillation of the electron kinetic energy
induced by a combined action of the ac electric field and the dc static electron potential. The oscillation amplitude increases with decreasing the frequency $\omega$ until the latter becomes comparable with $\tau_{\varepsilon}^{-1}$ whereupon 
the amplitude is stabilized by the energy relaxation processes.

The anisotropic oscillating correction $\delta f^{(EV)}_{{\bm k} \omega}$ describes the dynamic alignment of electron momenta in the $\bm k$-space. It has the form
\begin{eqnarray} \label{deltafEV}
\delta f^{(EV)}_{{\bm k}\,  \omega} =   {e v_0^2 \tau_{2\omega} \over 2 } \hspace{3 cm} \\ \times
\left[ (-f_0') \, \varepsilon_k \left( {\tau_{1 \omega}  \over \varepsilon_k}\right)' (E_{x} \cos{2\varphi_{\bm k}} + E_{y} \sin{2\varphi_{\bm k}}) {dV\over dx} \right. \nonumber \\ \left.
	+ f_0^{\prime\prime} \tau_{1 \omega} V(x ) \left( {d E_{x}\over dx} \cos{2\varphi_{\bm k}} + {d E_{y}\over dx} \sin{2\varphi_{\bm k}} \right) 	\right]\:. \nonumber
\end{eqnarray}
Here $\tau_{2\omega}^{-1} = \tau_2^{-1} - {\rm i} \omega$, where $\tau_2$ is the relaxation time of the second-order harmonics  of the distribution function proportional to $\cos{2\varphi_{\bm k}} = (k_x^2 - k^2_y)/k^2$ or $\sin{2\varphi_{\bm k}}= 2k_xk_y/k^2$. 

Substitution of the solution $f^{(EV)}_{{\bm k}\omega} = \left< f^{(EV)}_{k \omega} \right> + \delta f^{(EV)}_{{\bm k}\,  \omega}$ into Eq.~\eqref{j1} yields the ratchet current exactly in the form of Eq.~\eqref{phenom} where, see the details in Appendix,
\begin{subequations}
\begin{eqnarray} \label{chi_0}
\chi_0 &=& {e^3 v_0^2 \over 2 \pi \hbar^2} \left({\rm Re} S_1 - {{\rm Im} S_2} \right)\:, \\
\chi_L &=& \tilde{\chi}_L = -{e^3 v_0^2 \over 2 \pi \hbar^2}  {{\rm Im} S_2}\:, \label{chi_1}\\
\label{gamma}
\gamma &=& {e^3 v_0^2 \over 2 \pi \hbar^2} \left({{\rm Re} S_2} - {\rm Im} S_1 \right)\:.
\end{eqnarray}
\end{subequations}
The complex coefficients $S_{1,2}$ are defined by the following expressions 
\begin{eqnarray} \label{S}
S_1 &=& \varepsilon^3 \left( \tau_1 \over \varepsilon \right)'  \tau_{2\omega} \left( \tau_{1\omega} \over \varepsilon \right)' - {1 \over 2} \left[ \varepsilon^2 \left( \tau_1 \over \varepsilon \right)' \tau_{2\omega} \tau_{1\omega}  \right]'\:, \nonumber 
\\
S_2 &=&	{\left( \tau_1 \varepsilon \right)'  \left( \tau_{1 \omega} \varepsilon \right)' \over \omega \varepsilon} - {1 \over 2\omega} \left[ \left( \tau_1 \varepsilon \right)' \tau_{1\omega} \right]'\:, 
\end{eqnarray}
where one should set $\varepsilon=\varepsilon_{\rm F}$.

One can see that, for the second mechanism, the ratchet current reveals contributions both dependent and independent of the polarization. 

\begin{figure}[t]
\includegraphics[width=0.9\linewidth]{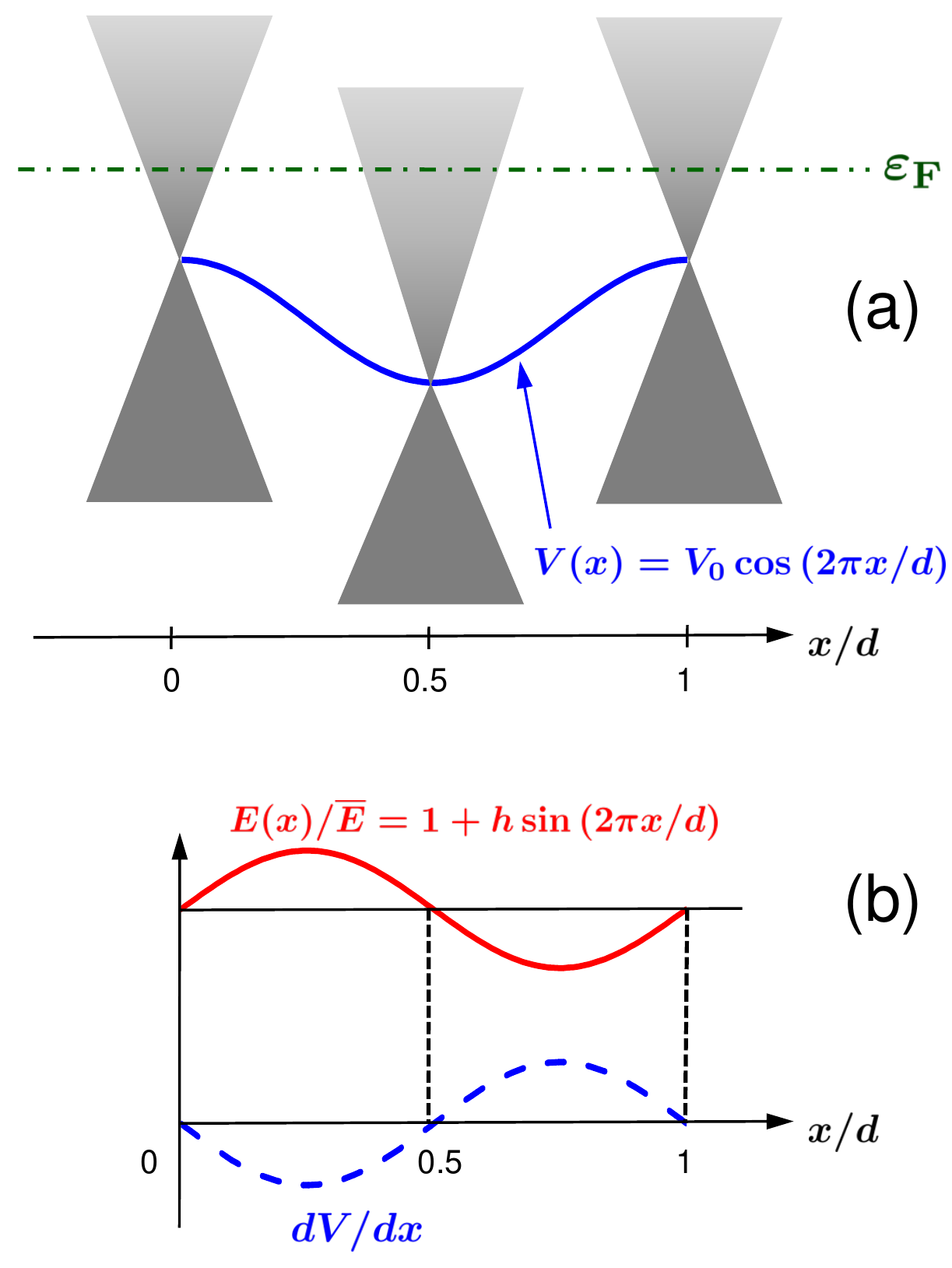}
\caption{(a) Schematic representation of the spatial variation of the Dirac point along the $x$ axis due to the built-in lateral potential $V(x)$ (solid curve), see Eq.~\eqref{energy}. Dashed-and-dotted horizontal line shows the equilibrium Fermi energy. (b) Spatial variation of the first derivative of the lateral potential and spatial modulation of the radiation electric field.}
\label{fig_illustration}
\end{figure}

In order to reveal the physical nature of the polarization-dependent ratchet currents we will consider in more detail one of the contributions to these currents, e.g., the contribution due to the first terms in the square brackets in Eq.~(\ref{f_EV_0}). Substituting this term to Eq.~(\ref{j1}), using the average (\ref{averageab}) and replacing $(- f'_0)$ by $\delta(\varepsilon - \varepsilon_F)$, see Appendix, we can reduce this contribution to the form
\begin{equation} 
%\label{onecontrib}
{\bm j} = \frac{e^2 v_0^2}{2 \varepsilon_{\rm F}} (\tau_1 \varepsilon)'_{\varepsilon=\varepsilon_{\rm F}} \overline{\delta N_{\omega}(x) {\bm E}^*(x)} + {\rm c.c.}
\end{equation}
Here $\delta N_{\omega}(x)$ is the electron density fluctuation related to the first term in Eq.~(\ref{f_EV_0}) and proportional to $E_x(x) (dV/dx)$:
\begin{equation} 
%\label{onecontrib}
\delta N_{\omega}(x) = \frac{{\rm i} e v_0^2}{2 \omega  \varepsilon_{\rm F}} (\tau_{1\omega} \varepsilon)'_{\varepsilon=\varepsilon_{\rm F}} g(\varepsilon_{\rm F}) E_x(x) \frac{dV}{dx} \:,
\end{equation}
where we introduced the density of states 
\begin{equation}
\label{DOS}
	g(\varepsilon) = \frac{2}{\pi} \frac{\varepsilon}{\hbar^2 v_0^2}\;.
\end{equation}
It is convenient to hold the further interpretation by using the illustration sketched in Fig.~\ref{fig_illustration}. The lateral potential is chosen in the form of $V_0 \cos{(2 \pi x/d)}$, as shown in Fig.~\ref{fig_illustration}a, and the phase of the electric-field amplitudes $E_{x,y}(x)$ is shifted by $\varphi_V - \varphi_E =\pi/2$ as follows: $E_{x,y}(x)/\overline{E}_{x,y} = 1 + h \sin{(2 \pi x/d)}$ as illustrated by Fig.~\ref{fig_illustration}b. In this case the products $\overline{E(x) (dV/dx)}$ and
\[
\overline{E^2(x) \frac{dV}{dx}} = 2 \overline{E} \, \overline{E(x) \frac{dV}{dx}}
%2 \overline{E} \overline{\left[ E(x) - \overline{E} \right] \frac{dV}{dx}}
\]
are nonzero. The time and space variation of $\delta N$ is represented by
\[
\delta N(x,t) \propto E_0 \frac{dV}{dx} \bigl( {\rm Im} S_2 \cos{\omega t} + {\rm Re} S_2 \sin{\omega t} \bigr)\:,
\]
where $E_0 = \overline{E}$ is the electric-field scalar amplitude, and $S_2$ is given by the first term in the second equation~(\ref{S}). For the circularly polarized light $E_x(t) \propto \cos{\omega t}, E_y(t) \propto \sin{\omega t}$ and, therefore, the time average of the product $\delta N(x,t) E_y(t)$ is proportional to ${\rm Re} S_2$. On the other hand, the $\tilde{\chi}_L$-related current in Eq.~(\ref{phenom}) is induced by the linearly polarized light with $E_x(t), E_y(t) \propto \cos{\omega t}$ resulting in $\tilde{\chi}_L \propto {\rm Im} S_2$, in agreement to Eqs.~(\ref{S}). The other terms in these equations are obtained in the similar way by taking into account the explicit expressions for $\left< f^{(EV)}_{k \omega} \right>$ and $\delta f^{(EV)}_{{\bm k}\,  \omega}$.

\subsection{Quantum-well structures}

In this subsection we extend the theory of Refs.~\onlinecite{Review_JETP_Lett,Olbrich_PRL_09,Olbrich_PRB_11} to consider the degenerate statistics of the two-dimensional electron gas in heterostructures and take into account the possible difference between the relaxation times $\tau_1$ and $\tau_2$ and their energy dependence.

Solution of the kinetic Eq.~\eqref{EV} shows that Eqs.~\eqref{f_EV_0} and~\eqref{deltafEV} are also valid for the QW structures provided  $v_0^2$ and 
$(\tau_{1 \omega}/\varepsilon_k)'$ are replaced by the squared Fermi velocity $v_{\rm F}^2={2\varepsilon_{\rm F}/m}$ and   $\tau_{1 \omega}'/\varepsilon_k$, respectively. Furthermore, for the QW structures we also obtain Eqs.~\eqref{chi_0}--\eqref{gamma} with $v_{\rm F}^2$ instead of $v_0^2$ and $S_{1,2}$ given by
\begin{eqnarray}
\label{S_QW}
S_{1,{\rm QW}} &=& \varepsilon \tau_1' \tau_{2\omega} \tau_{1\omega}' - {1 \over 2\varepsilon} \left( \varepsilon^2 \tau_1' \tau_{2\omega} \tau_{1\omega}  \right)'\:,  \\
S_{2,{\rm QW}} &=&	{\left( \tau_1 \varepsilon \right)'  \left( \tau_{1 \omega} \varepsilon \right)' \over \omega \varepsilon} - {1 \over 2\omega \varepsilon} \left[ \left( \tau_1 \varepsilon \right)' \tau_{1\omega} \varepsilon \right]'\:. \nonumber  
\end{eqnarray}

We note that for a quadratic energy dispersion $\varepsilon_k=\hbar^2k^2/(2m)$ and energy-independent time $\tau_1 \equiv \tau_{\rm tr}$, the partial derivative 
$
{\partial}\bigl(\tau_1 v_{\alpha}  \bigr)/{\partial k_{\beta}}  = \delta_{\alpha \beta} {\hbar \tau_{\rm tr}}/{m}
$ 
is independent of ${\bm k}$, and Eq.~(\ref{j1}) for the ratchet current takes the form~\cite{Review_JETP_Lett}
\[
{\bm j} = {2 e^2\tau_{\rm tr}\over m} \, {\rm Re}\left[ \overline{\delta N_\omega(x) {\bm E}^*(x)}\right]\:,
\]
where $\delta N_\omega(x)= 2 \sum\limits_{\bm k} f^{(EV)}_{{\bm k} \omega}$ is the second-order correction to the electron density.

\section{Discussion} \label{sec:discussion}

\begin{figure*}[t]
\includegraphics[width=0.4\linewidth]{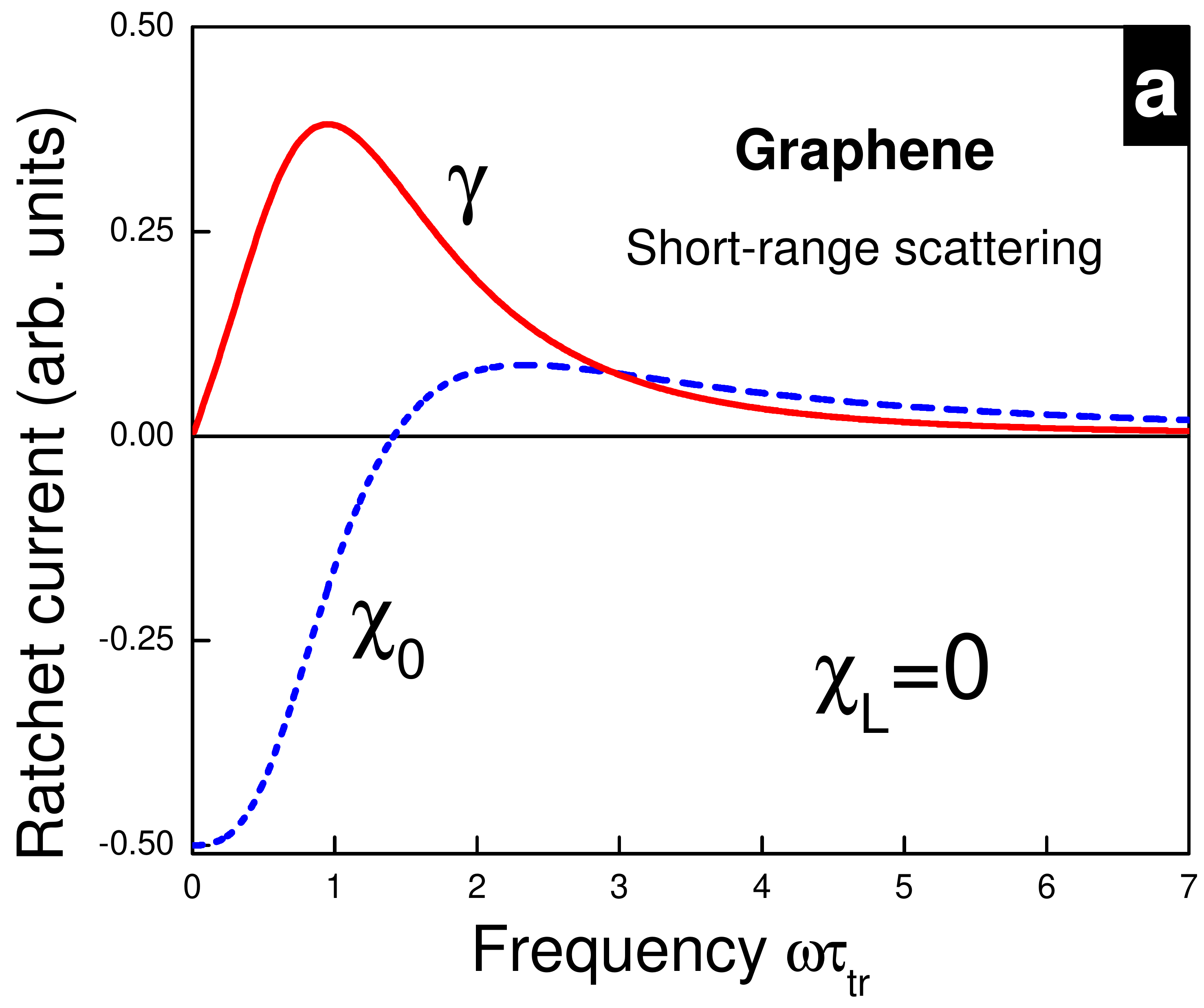}
\qquad
\includegraphics[width=0.4\linewidth]{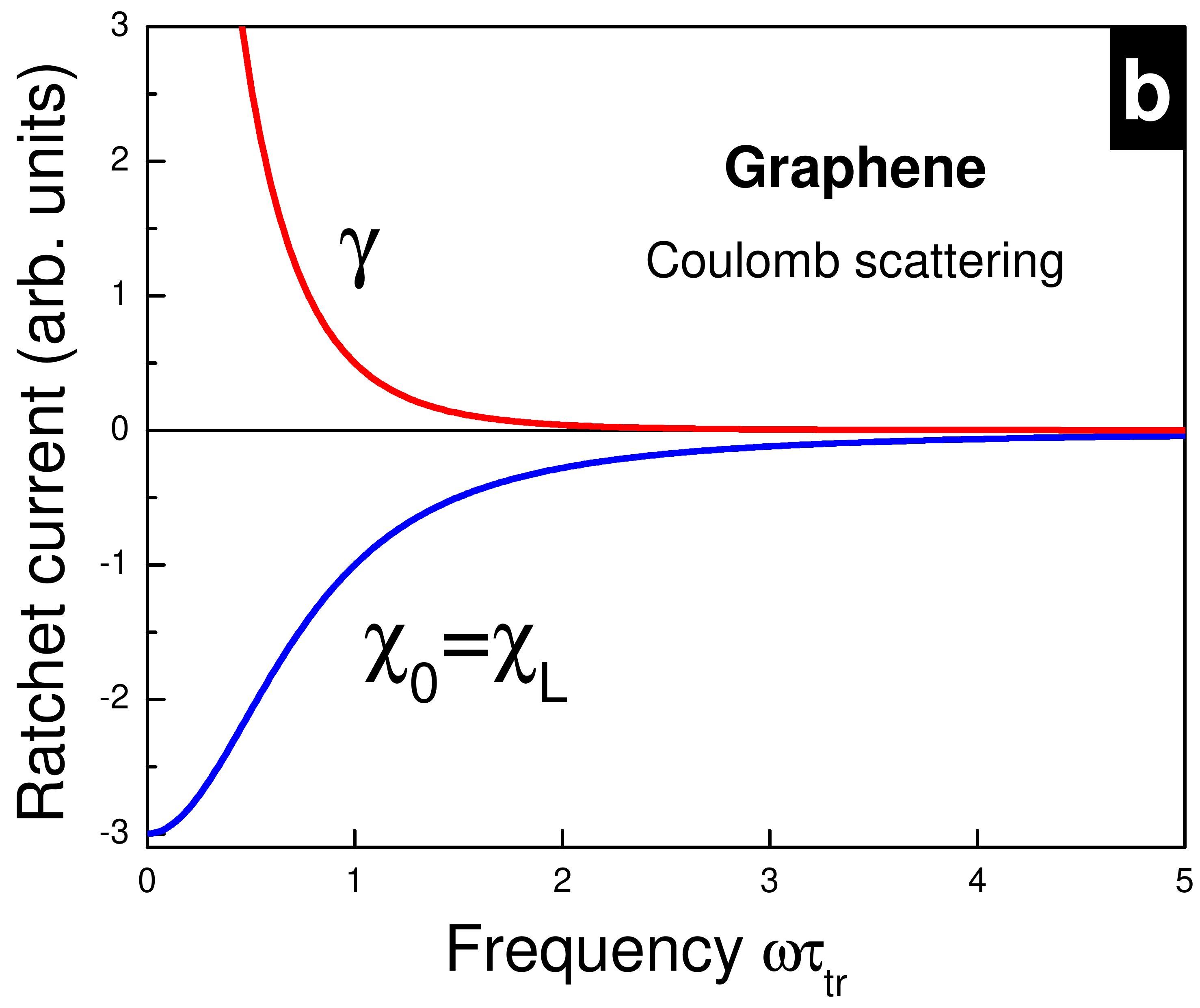}
\caption{Frequency dependencies  of ratchet currents in graphene for scattering by short-range defects~(a) and Coulomb impurities~(b). 
%The linear-polarization-dependent contribution $\chi_L$ is absent for short-range scattering while for the Coulomb scattering it is equal to the polarization-independent term $\chi_0$. 
%The helicity-dependent contribution $\gamma$ is present in both cases and equals to zero at $\omega \to 0$, but for Coulomb scattering this takes place at $\omega \sim \tau_\varepsilon^{-1}\ll \tau_{\rm tr}^{-1}$.
}
\label{fig1}
\end{figure*}

\begin{figure*}[t]
\includegraphics[width=0.4\linewidth]{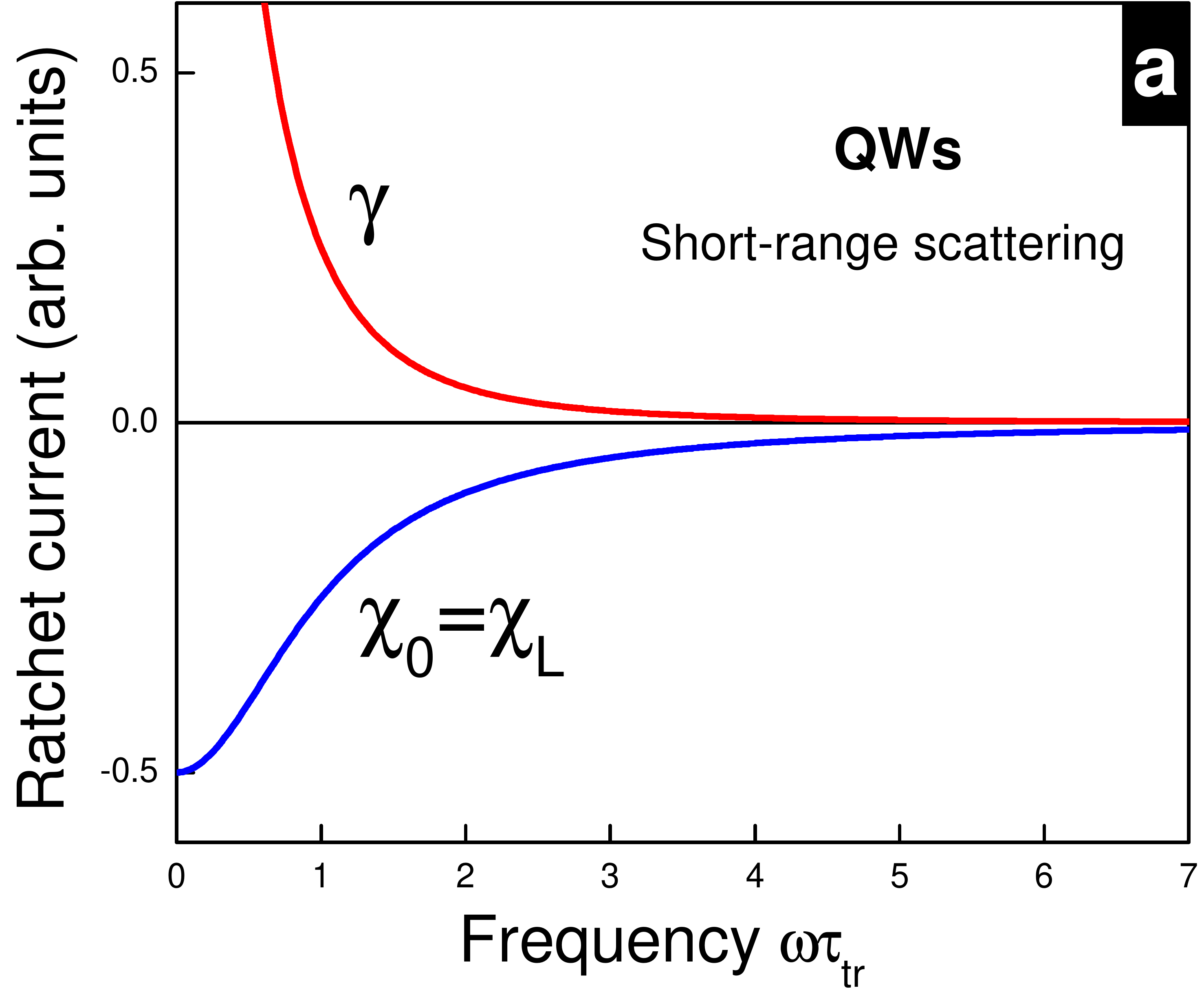}
\qquad
\includegraphics[width=0.4\linewidth]{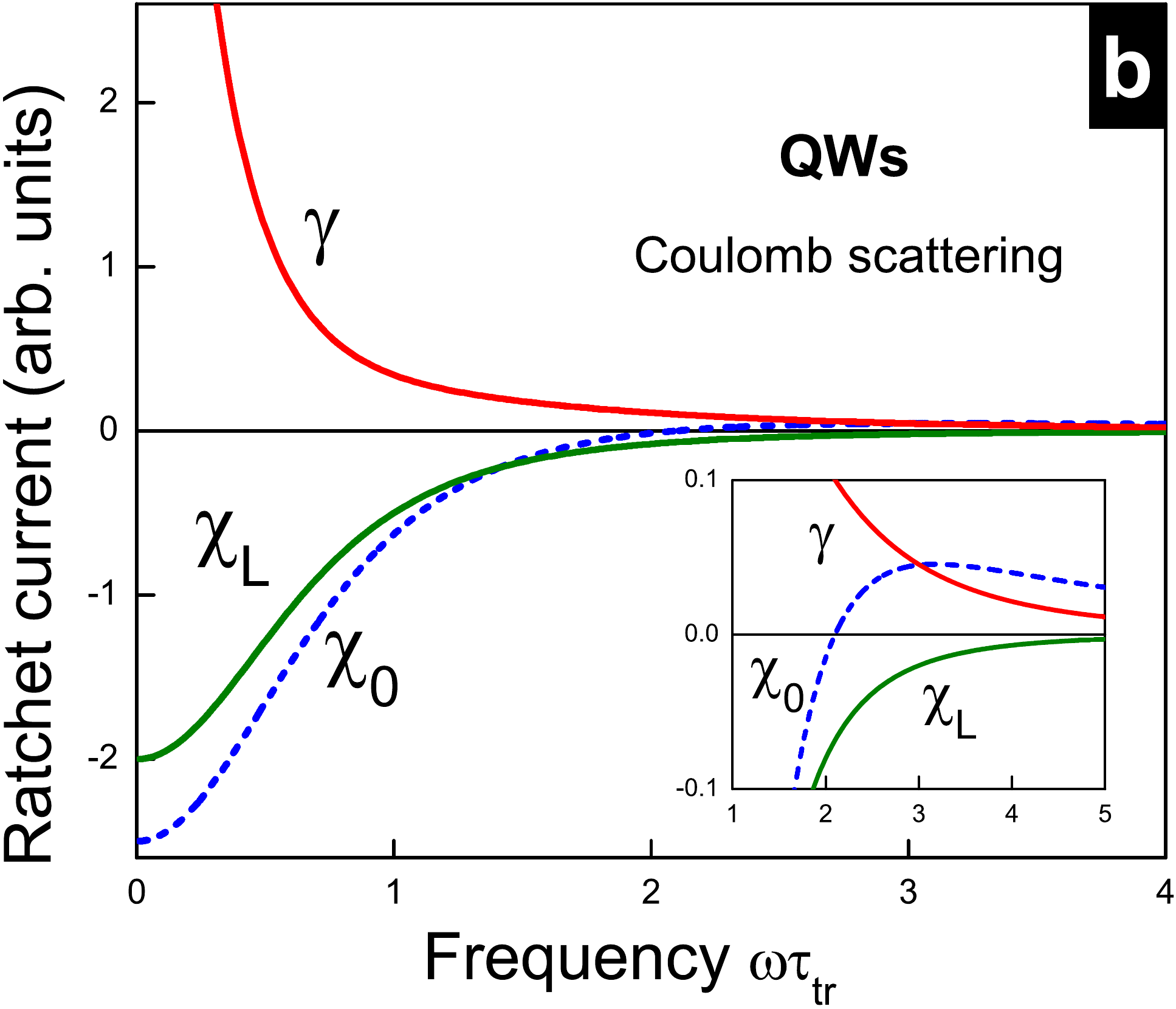}
\caption{
%Excitation spectra 
Frequency dependencies of ratchet currents in QW structures at two types of elastic scattering: by short-range defects (a) and by Coulomb impurities (b). 
%In both cases $\gamma \to 0$ at $\omega\tau_{\rm tr} \sim \tau_{\rm tr}/\tau_\varepsilon \ll 1$. 
Inset shows the frequency range where $\chi_0$ changes its sign.}
\label{fig_QW}
\end{figure*}

We calculate the ratchet current excitation spectrum for two types of elastic scattering actual for graphene. For scattering by short-range defects 
%in graphene 
one has
\[\tau_1=\tau_{\rm tr} {\varepsilon_{\rm F}\over \varepsilon}, 
\qquad
\tau_2 = {\tau_1 \over 2},
\] 
and Eqs.~\eqref{S} yield $\chi_L=0$ for this case. In contrast, for scattering by Coulomb impurities, when
\[\tau_1=\tau_{\rm tr} {\varepsilon \over \varepsilon_{\rm F}}, 
\qquad
\tau_2 = 3 \tau_1,
\] 
$S_1=0$, and $\chi_L=\chi_0$. This means that $x$ component of the current is generated in this case only by $x$-polarized radiation: $j_x \propto |e_x|^2$.

In Fig.~\ref{fig1} we plot the ratchet current frequency dependence 
%excitation spectrum 
for both types of scattering.
It can be seen that the coefficients $\chi_{0,L}$ and $\gamma$  have complex non-monotonous behavior. In the static limit, $\omega \to 0$, $\chi_0$ remains finite while the circular ratchet current is absent, $\gamma \to 0$. This is correct because  helicity-dependent effects can not be present for static electric field. For Coulomb scattering $\gamma$ also tends to zero but this occurs at $\omega \sim \tau_\varepsilon^{-1} \ll \tau_{\rm tr}^{-1}$ as discussed in the paragraph below Eq.~\eqref{f_EV_0}.

Now we turn to the QW structures. 
For scattering by short-range defects in QWs when
\[
\tau_1=\tau_2 = \tau_{\rm tr},
\] 
Eqs.~\eqref{S_QW} yield
\[
\chi_0=\chi_L=\tilde{\chi}_L = -\omega\tau_{\rm tr} \, \gamma,
\quad
	\gamma = {e^3 \over 2 \pi \hbar^2 m \omega} {\tau_{\rm tr}^2 \over 1 + (\omega\tau_{\rm tr})^2}.
\]
Again, we obtain generation of $j_x$ at radiation polarization along the $x$ axis only, but, in contrast to graphene, this takes place for short-range scattering.
Comparing this expression with the result for Boltzmann statistics valid at room temperature $T_{\rm room}$,~\cite{Review_JETP_Lett} we get
\[
{\gamma(T=0) \over \gamma(T_{\rm room})} \sim {T_{\rm room} \over \varepsilon_{\rm F}}.
\]
This estimation implies that the helicity-dependent ratchet current 
shows no remarkable variation with temperature.
%does not change strongly with temperature.

For scattering by Coulomb impurities in QWs, when
\[\tau_1=\tau_{\rm tr} {\varepsilon \over \varepsilon_{\rm F}}, 
\qquad
\tau_2 = \tau_1/2,
\] 
we obtain $\chi_0 \neq \chi_L$. 
Figure~\ref{fig_QW} 
shows the ratchet current in QW structures.
One can see that, for Coulomb scattering, 
polarization-dependent ratchet currents are sign-constant, while $\chi_0$ has a maximum and changes its sign at $\omega\tau_{\rm tr} \approx 2$, see inset in Fig.~\ref{fig_QW}.

The Seebeck contribution to the ratchet current is absent in graphene for both considered types of elastic scattering, cf. Eq.~\eqref{chi_Seebeck}.  
Nonzero Seebeck ratchet current in graphene appears, e.g., at scattering by screened Coulomb potential.

The Seebeck ratchet current 
is also absent for short-range scattering in QWs, cf. Eq.~\eqref{chi_Seebeck_QW}. In contrast,  for scattering by Coulomb impurities $\chi_{0,QW}^{\rm S}$ is nonzero. Its ratio to the elastic-scattering contribution can  be estimated as 
\[
{\chi_{0,QW}^{\rm S} \over \chi_{0,QW}} \sim \pi^2 {\tau_\varepsilon \over \tau_{\rm tr}}{T \over \varepsilon_{\rm F}}.
\] 
At the low temperature $T \approx 4$~K, the energy relaxation time in QWs 
$\tau_\varepsilon \sim 1$~ns~\cite{en_rel} while the transport scattering time $\tau_{\rm tr}\sim 1$~ps. Therefore the Seebeck contribution to the polarization-independent ratchet current dominates in QW structures at low temperatures for scattering by the smooth Coulomb potential.

\section{Summary} \label{sec:concl}

To summarize, radiation induced electric currents in graphene with a spatially periodic noncentrosymmetric lateral potential are studied theoretically.
The ratchet current is shown to consist of polarization-independent contribution and the contribution sensitive to linear and circular polarization of radiation.
Two microscopic mechanisms of the polarization-independent ratchet current are considered and compared: the Seebeck contribution generated in the course of energy relaxation and the current controlled by elastic scattering processes.
We demonstrate that the ratchet current excitation spectrum strongly depends on the type of elastic scattering.  Two realistic mechanisms of electron scattering in graphene are analyzed.
For a short-range potential, there are polarization-independent and helicity-dependent currents, while the linear polarization leads to no ratchet current. For the Coulomb scattering, the linearly-polarized radiation generates the ratchet current only for the polarization vector parallel to the lateral-potential modulation direction. The Seebeck ratchet current is shown to vanish for both types of elastic scattering in graphene.
For comparison, we have analyzed the ratchet effect in QW structures with a lateral superlattice and degenerate electron gas and demonstrated the  polarization-dependent effects as well as the Seebeck ratchet current for the Coulomb scattering.
These results show that ratchet current measurements allow one to identify a dominant mechanism of elastic scattering in graphene.

\acknowledgments{We thank S.D. Ganichev for stimulating discussions. The work was supported by RFBR, President grant for young scientists, Program ``Leading Scientific Schools'' (\#5442.2012.2), and EU program ``POLAPHEN''.}

\appendix
\section{Derivation of equations for $S_1$ and $S_2$}
In order to calculate explicitly the current (\ref{j1}) for graphene one can apply the identity 
\begin{equation} \label{alphabeta}
 \frac{\partial \bigl(\tau_1 v_{\beta}  \bigr)}{\partial k_{\alpha}} = \hbar v_0^2 \left[ \frac{\tau_1}{\varepsilon} \delta_{\alpha \beta} +  \varepsilon \left( \frac{\tau_1}{\varepsilon} \right)' \frac{k_{\alpha} k_{\beta}}{k^2}\right]\:.
\end{equation}

The summation over ${\bm k}$ in (\ref{j1}) is performed in two stages, firstly, by averaging over the directions of the wavevector ${\bm k}$ and, secondly, by integration over the modulus $k \equiv |{\bm k}|$ or, equivalently, over the energy
$\varepsilon \equiv \varepsilon_k = \hbar v_0 k$. 
For finding the contribution of $\left<  f^{(EV)}_{k \omega} \right>$ to the ratchet current, it suffices to use the average
\begin{equation} \label{averageab}
\left< \frac{\partial \bigl(\tau_1 v_{\beta}  \bigr)}{\partial k_{\alpha}} \right> = \frac{\hbar v_0^2}{2}~\frac{(\tau_1 \varepsilon)'}{\varepsilon}~ \delta_{\alpha \beta}\:.
\end{equation}
Since the anisotropic correction $\delta f^{(EV)}_{k \omega}$ is a linear function of $\cos{2 \varphi_{\bm k}}$ and $\sin{2 \varphi_{\bm k}}$, its contribution to the ratchet current ${\bm j}$ is found by using the identities
\begin{eqnarray}
\left< \cos{ 2 \varphi_{\bm k} } \frac{\partial \bigl( \tau_1 v_x  \bigr) }{ \partial k_x} \right>
&=& - \left< \cos{2 \varphi_{\bm k}} \frac{\partial \bigl(\tau_1 v_y  \bigr)}{\partial k_y} \right>
\nonumber\\ = \left< \sin{2 \varphi_{\bm k}} \frac{\partial \bigl(\tau_1 v_x  \bigr)}{\partial k_y} \right>
&=& \left< \sin{2 \varphi_{\bm k}} \frac{\partial \bigl(\tau_1 v_y  \bigr)}{\partial k_x} \right> \nonumber\\
&=&
%\hspace{2 cm}= 
\frac{\hbar v_0^2}{4}~\varepsilon \left( \frac{\tau_1}{\varepsilon} \right)'\:.
\end{eqnarray}

The calculation at the second stage, integration over the energy, is simplified by the assumption of the degenerate 
statistics in which case the the first derivative $f'_0$ can be replaced by the delta-function $- \delta(\varepsilon - \varepsilon_F)$. The terms in $f^{(EV)}_{\bm k \omega}$ proportional to the second derivative $f''_0$ are treated by using the identity
\begin{equation} \label{secondderiv}
4 \sum_{\bm k} f''_0(\varepsilon_k) F(\varepsilon_k) = \left[ g(\varepsilon) F(\varepsilon)\right]'_{\varepsilon = \varepsilon_{\rm F}}\:,
\end{equation}
where $F(\varepsilon)$ is an arbitrary smooth function of $\varepsilon$ and $g(\varepsilon)$ is the electron density of states given by Eq.~\eqref{DOS}.
%$(2/\pi)\varepsilon/(\hbar v_0)^2$.

\end{document}